\documentclass[12pt,preprint]{aastex}

\begin{document}

\slugcomment{}
\shorttitle{PDR Model Mapping of M17-SW}
\shortauthors{Sheffer \& Wolfire}

\title{PDR Model Mapping of Obscured H$_2$ Emission and the Line-of-Sight Structure of M17-SW}

\author{Y. Sheffer and M. G. Wolfire}

\affil{Department of Astronomy, University of Maryland, College Park, MD 20742, USA}

\begin{abstract}

We observed H$_2$ line emission with \textit{Spitzer}-IRS toward M17-SW and
modeled the data with our PDR code.
Derived gas density values of up to few times 10$^7$ cm$^{-3}$ indicate that
H$_2$ emission originates in high-density clumps.
We discover that the PDR code can be utilized to map the amount of intervening
extinction obscuring the H$_2$ emission layers, and thus we obtain the radial
profile of $A_V$ relative to the central ionizing cluster NGC 6618.
The extinction has a positive radial gradient, varying between 15---47 mag
over the projected distance of 0.9---2.5 pc from the primary ionizer, CEN 1.
These high extinction values are in good agreement with previous studies
of $A_V$ toward stellar targets in M17-SW.
The ratio of data to PDR model values is used to infer the global line-of-sight
structure of the PDR surface, which is revealed to resemble a concave surface
relative to NGC 6618.
Such a configuration confirms that this PDR can be described as a
bowl-shaped boundary of the central \ion{H}{2} region in M17.
The derived structure and physical conditions are important for interpreting
the fine-structure and rotational line emission from the PDR.

\end{abstract}

\keywords{infrared: ISM --- ISM: clouds --- ISM: individual objects (M17) --- ISM: molecules --- photon-dominated region (PDR) --- open clusters and associations: individual (NGC 6618)}

\section{Introduction}

The star forming region M17 is notable for its asymmetry in terms of its
appearance in different wavelength regimes.
Optically, it is visible as a nebula with a prominent northern bar marking
the location of an ionization front, which suffers only a low level of
optical extinction \citep[][and references therein]{Felli1984}.
On the other hand, radio maps have long revealed the presence of the
southern bar, part of the M17-SW region,
which has no obvious optical counterpart, and is thus understood to suffer
a much higher level of extinction along our line of sight (LOS).
Also noteworthy is the high-extinction patch situated in front of a number
of O-type stellar members belonging to the ionizing cluster NGC 6618,
which from our direction can be found between the two ionization bars
\citep[e.g.,][]{Dickel1968,Beetz1976,Povich2007}.

Given that the southern bar includes an ionization front and a PDR that are
both obscured in optical wavelength, it is best studied in the IR
and radio regimes.
Previous IR and radio studies of M17-SW have described it as almost edge-on
PDR \citep{Stutzki1988}, or as a bowl carved into the molecular cloud
\citep{Meixner1992,Brogan2001,Pellegrini2006}, and have concluded that it is
comprised of a clumpy medium \citep{Stutzki1990,Meixner1992}.

For this study we obtained \textit{Spitzer} spectroscopy of rotational emission
lines of H$_2$ toward M17-SW in order to model the data with our PDR code.
Specifically, we performed model mapping of the gas density over the field
of view, and inferred the LOS configuration of the PDR layer
\citep[see][]{Sheffer2011}.
Owing to the presence of extinction obscuring the interface between the
\ion{H}{2} region and the molecular cloud, we also included $A_V$ in model
mapping as means of dereddening the H$_2$ line intensities.
We shall present our results as a function of the radial distance from CEN 1,
the primary O4 member of NGC 6618.

\section{\textit{Spitzer}-IRS Data and Analysis}

Our \textit{Spitzer} observations of M17-SW belong to programs P03697
(SH and LH data) and P30295 (SL data).
Data acquisitions for P03697 were executed on three dates between 2004 Oct 02
and 2005 Apr 23, and those for P30295 on the two dates 2007 Sep 30 and Oct 12.
The target area was covered by multiple AORs: 11 for SH,
and 4 each for LH and SL data, see Table 1.
Following re-gridding onto the LH pixel frame, the area available for full
analysis based on five emission lines includes 480 LH pixels.
Each LH pixel is 4\farcs46 wide, corresponding to 8900 AU, or 0.043 pc
at a distance of 2.0 kpc \citep[$\pm7\%$ precision, see][]{Xu2011}.

\begin{figure}
\epsscale{1.3}
\plottwo{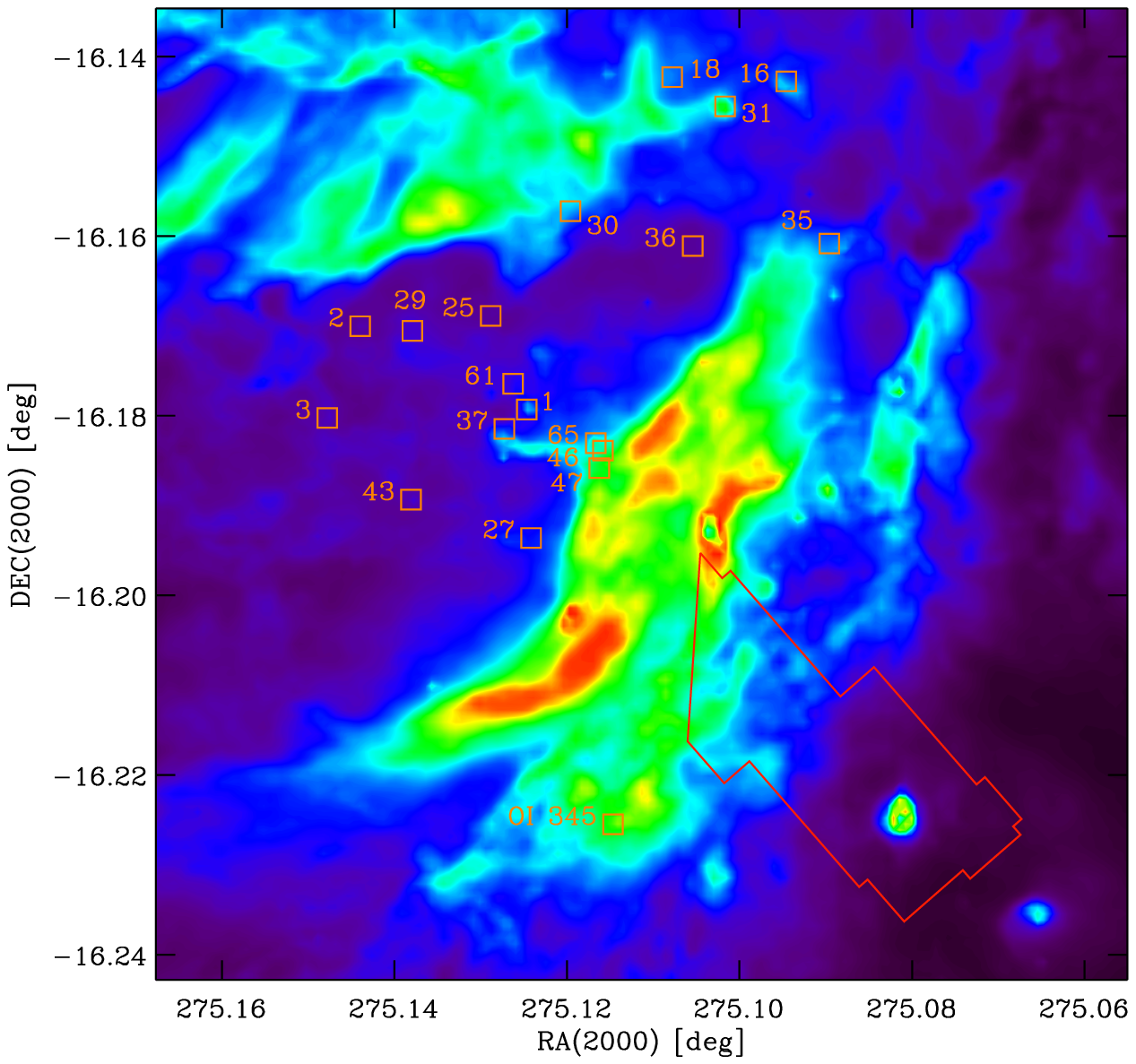}{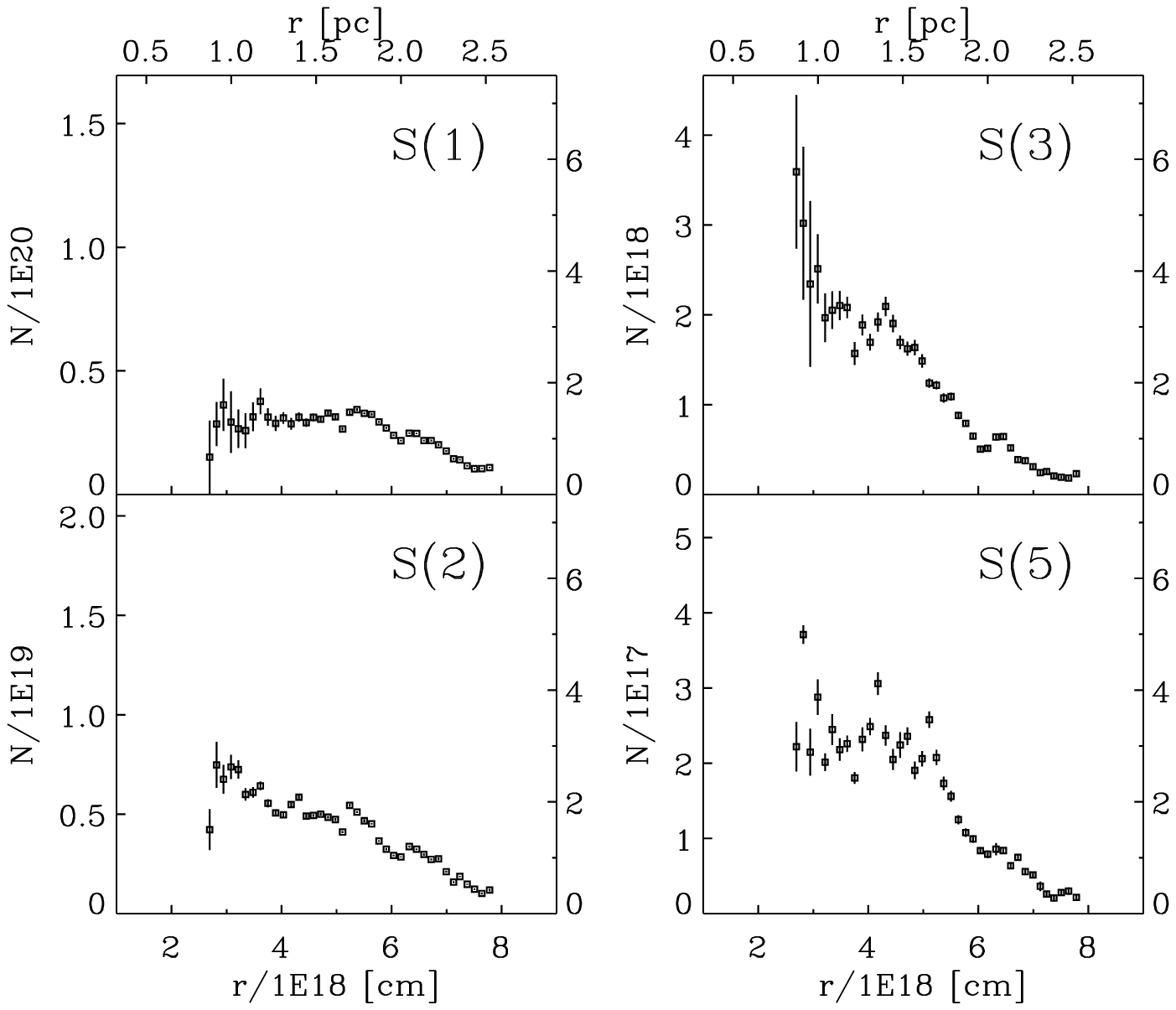}
\caption{Left panel shows the area of overlap for all IRS modules
(red irregular outline) overlaying the IRAC channel 4 image of M17-SW.
The positions of 19 O stars are indicated by squares, along with their CEN
numbers \citep{Chini1980}.
The IR-bright point source inside our field is known as the KW object
\citep{Kleinmann1973}.
Right panels present observed intensity values for four H$_2$ emission lines,
using a common intensity scale (right ordinate values, in
units of 10$^{-4}$ erg s$^{-1}$ cm$^{-2}$ sr$^{-1}$).
Error bars on intensity values show the $\pm2\sigma$ uncertainties.
Each intensity plot is converted into column density values, as given by the
left ordinate scales in cm$^{-2}$.}
\label{fig:fig1}
\end{figure}

We employed version 1.7 of CUBISM \citep{Smith2007} for the reduction of
observations and the construction of data cubes therefrom,
assuring a match with the \textit{Spitzer} Science Center pipeline version
S18.7.0.
Left panel of Figure~\ref{fig:fig1} shows the proper celestial location and
orientation
of the area of intersection of all modules over an 8 $\micron$ image from
\textit{Spitzer}/IRAC.

Four pure-rotational emission lines of H$_2$ were detected and mapped toward
the target: S(1) and S(2) at 17.03 and 12.28 $\micron$, respectively, from
the SH module, and S(3) and S(5)
\clearpage
\clearpage
\noindent at 9.66 and 6.91 $\micron$, respectively,
from orders 1 and 2, respectively, of the SL module.
Owing to lack of reliable signal from S(0) at 28.22 $\micron$,
albeit located within the wavelength interval of coverage of the LH module,
our PDR modeling did not include the S(0) line.
Emission line maps were constructed by using in-house IDL procedures to fit
line profiles and to derive integrated line intensities that included continuum
removal following its fitting by a low-order polynomial.
The presence in these spectra of emission lines of ionized atomic species
appreciably stronger than the emission lines from H$_2$ necessitated an
additional step of deblending S(5) and [\ion{Ar}{2}] at 6.98 $\micron$.
Such a challenge did not arise in the analysis of similar data toward
NGC 2023-South, where lines of ionized species were either weak or undetectable
\citep{Sheffer2011}.
We note that the spectral co-presence of \ion{H}{2} region and PDR emission
lines shows that both regions are sampled along the LOS toward M17-SW,
thus indicating that their configuration cannot be strictly edge-on.

No attempt was made to correct for zodiacal background emission owing to the
bright nature of this target and the insensitivity of
continuum-subtracted emission lines to such uniform contribution.
Based on the discussion in \cite{Sheffer2011}, data from the LH and SH modules
were corrected by a factor of 0.84 in order to obtain calibration match with
the SL module flux values.

In order to compare emission data with PDR models, line intensities were
converted to column density via $N_J=4\pi I_J/A_J\Delta E_J$ cm$^{-2}$,
where $N_J, I_J, A_J$, and $\Delta E_J$ stand for the column density, emission
intensity, Einstein A-coefficient, and transition energy for each
rotational upper level $J$.
This conversion is linear owing to insignificant self-absorption of these
quadrupole transitions.
The four right panels of Figure~\ref{fig:fig1} show the radial variation of
H$_2$ emission lines with distance from CEN 1, following data averaging
along the orthogonal direction.
Both intensity and column density scales are provided.

\section{PDR Modeling of H$_2$ Emission}

The two primary PDR parameters are $n_{\rm H}$, the total hydrogen number
density, and $G_0$, the ratio of the incident 6---13.6 eV far-ultraviolet (FUV)
flux over the Habing flux of $1.6\times10^{-3}$ erg s$^{-1}$ cm$^{-2}$
\citep{Habing1968}.
We employed the \cite{Kaufman2006} PDR code to generate a ($n_{\rm H}$, $G_0$)
grid of 950 normal models, where `normal' means that
the incident radiation field is normal to the PDR surface ($\phi=0\degr$).
Model output consists of $N_J$(H$_2$) values following an integration
along the normal as well, with grid step being 0.1 dex.

Model mapping was performed with the parameter $G_0$ constrained
to predicted values.
This was motivated by the availability of lists of O-type stars residing
in and around NGC 6618 \citep[e.g.,][]{Broos2007,Povich2009}, as well as by the
knowledge of precise (albeit projected) linear distances between each star
and each mapped data pixel.
Thus $G_0$ was obtained by summing the FUV flux of 19 O-type stars listed in
\cite{Hoffmeister2008} and using
\begin{equation}
G_0^{\rm pr}=\frac{850}{D^2}\sum_i\frac{L_i^{\rm FUV}}{\theta_i^2},
\end{equation}
where the superscript `pr' may stand for `predicted' or `projected,'
$D$ is the distance to M17-SW in kpc, $L_i^{\rm FUV}$ is the FUV
luminosity of stellar radiator $i$ in solar units \citep{Parravano2003},
and $\theta_i$ is the angular separation between star $i$ and any map pixel
in seconds of arc.
The left panel of Figure~\ref{fig:fig2} shows that the values of $G_0^{\rm pr}$
range over $\sim$(1---8)$\times10^4$.

We perform a search for the smallest
root mean square deviation (RMSD) of the differences
in dex between modeled and observed $N_J$ values, thus yielding
values for the ratio $f_{\rm eff}=$ data/model.
This ratio may be decomposed into two $\geq1$ and two $\leq1$ factors,
$f_{\rm eff}=f_{\rm P}f_\theta\times f_\phi f_{\rm B}$,
where $f_{\rm P}$ is the number of PDRs along the LOS and here assumed
to be 1, $f_\theta=1/\cos(\theta)$ accounts for limb brightening owing to
inclination angle $\theta$, $f_\phi=\cos(\phi)$ accounts for the angle of
incidence of the radiation field on the PDR, and $f_{\rm B}$ is the beam area
filling factor, also assumed to be 1.
This assumption is consistent with the $\geq0.1$ pc sizes of
both observationally-derived C$^{18}$O clumps toward M17-SW
\citep{Stutzki1990} and PDR-modelled clumps with $n_{\rm H}=10^7$ cm$^{-3}$
\citep{Meixner1993}, as well as with the $\leq0.04$ pc (at $D=2.0$ kpc)
beam widths of the SH and SL modules of \textit{Spitzer}-IRS.
Owing to the dependence of the two angular factors, $f_\theta$ and $f_\phi$,
on the local orientation of the cloud surface \citep{Sheffer2011},
we shall employ them to infer clues about the LOS variations
of the PDR over the field of view, see $\S$5.

\section{PDR Modeling with $A_V$ as a Free Parameter}

It is customary to de-redden the observed $I_J$ values by the known or assumed
value of extinction, $A_V$, prior to comparing $A_V$-corrected
$N_J$ values with model output.
Here we employ the \cite{Mathis1990} reddening law.
Our initial PDR mapping runs employed a global (or spatially-invariant)
correction by $A_V=8$ mag, approximating
the average value toward the central members of NGC 6618
\citep{Hanson1997,Povich2007}.
However, some of the OB stars in the M17 region appear to show appreciably
higher extinction values, with $A_V\geq15$ mag
\citep{Tokunaga1979,Chini1980,Hanson1997}.
Follow-up modeling with $A_V=15$ mag returned fits with RMSD values smaller by
factors of 1.3---1.5, whereas further reductions by factors of 1.3---1.8 were
achieved by fits employing a global $A_V>20$ mag.
Inevitably, we included $A_V$ as a free parameter that was allowed to
\clearpage
\begin{figure}
\epsscale{1.4}
\plottwo{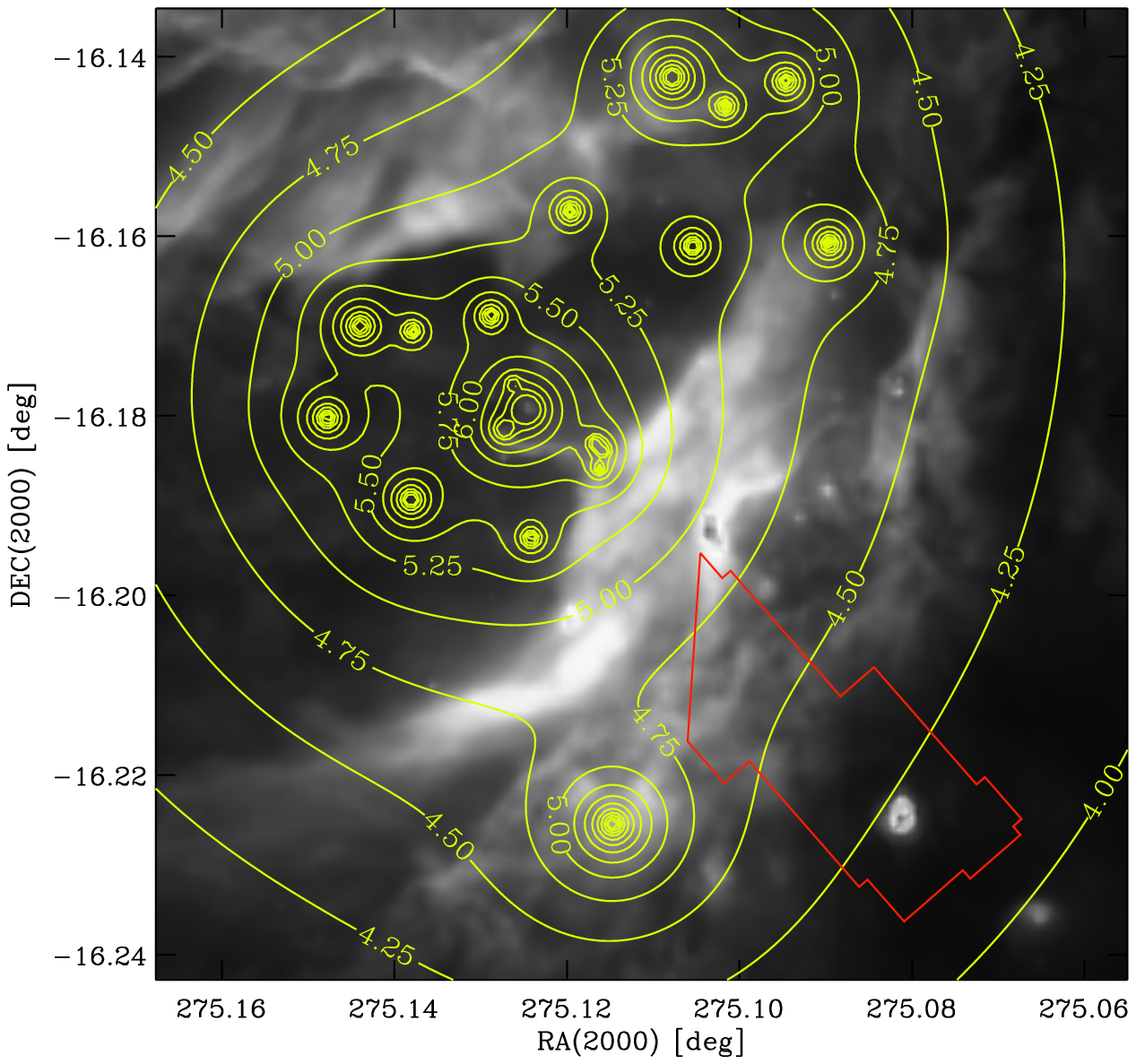}{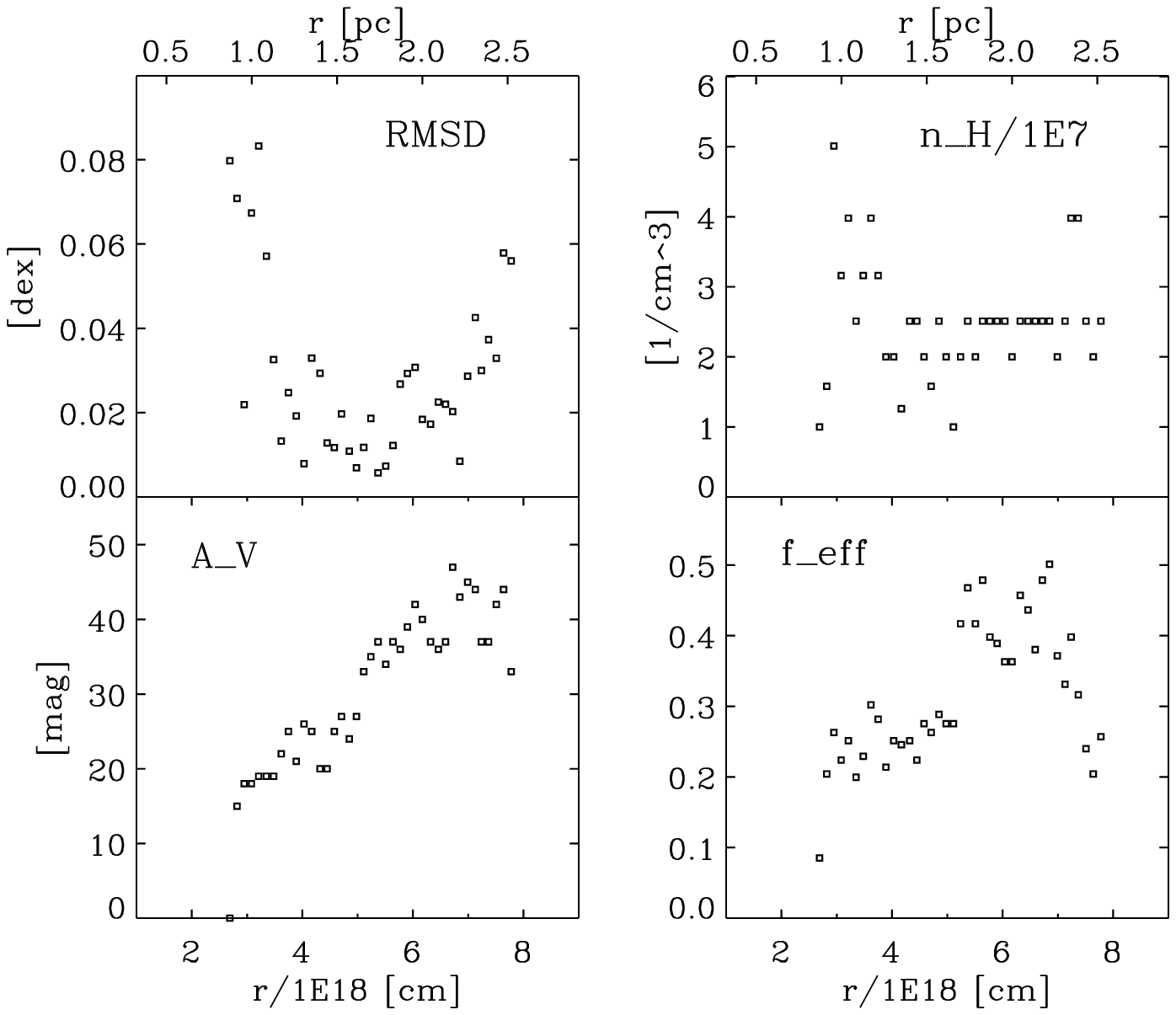}
\caption{Left panel shows log $G_0^{\rm pr}$ values from Eq. 1, based on
$D=2.0$ kpc.
Stellar FUV luminosities for 19 O-type members of NGC 6618 are summed,
including luminosity values corrected by $\times4$ for CEN 1, and
by $\times2$ for CEN 3, CEN 18, and CEN 37, owing to stellar multiplicity
\citep{Hoffmeister2008}.
Right panels show modeled physical parameters as a function of distance from
CEN 1.
Top panels show the RMSD of the fits, with a mean of 0.03 dex (7\%), and
$n_{\rm H}$, with a mean of $2.5\times10^7$ cm$^{-3}$.
Lower panels provide modeled values of $A_V$ toward M17-SW, and of
$f_{\rm eff}$, the ratio of data to modeled H$_2$ line intensities.}
\label{fig:fig2}
\end{figure}
\clearpage
\noindent vary over the entire field of view.
Superficially, both $f_{\rm eff}$ and $A_V$ corrections affect the ratios
of data to model intensities.
However, whereas the former applies the same factor to all emission lines,
the latter has different values for different emission lines.
Thus $A_V$ affects the ratios between emission lines from different
$J$ levels, unlike $f_{\rm eff}$.

The right panels of Figure~\ref{fig:fig2} present our PDR model mapping results,
with a mean RMSD of 0.029, or $\sim$7\%.
Values of $n_{\rm H}$ are found to be mostly between (2.0---2.5) $\times10^7$
cm$^{-3}$, or $\sim100$ times higher than the density modeled for the
Southern Ridge (SR) in NGC 2023 \citep{Sheffer2011}.
Such values are consistent with H$_2$ emission production in high-density
clumps immersed in an interclump gas of density lower by 2 or 3 orders
of magnitude \citep{Meixner1993}.
The (unshown) fixed parameter $G_0$ follows the values of $G_0^{\rm pr}$
from Eq. 1 by design.
The next panel presents model output for $A_V$, which to our knowledge
is the first attempt to derive extinction values from PDR modeling.
The mean of visual extinction preferred by the models is 30 mag, about 2---4
times as high as the initially presumed values of 8 and 15 mag.
Values of $A_V$, which range over 15---47 mag, are clearly increasing away
from NGC 6618 with a gradient of 21 $\pm$ 2 mag pc$^{-1}$.
Such a positive gradient is not unexpected: the structure of M17-SW includes a
sequence of \ion{H}{2} region, a PDR, and a dense molecular cloud along
the same radial direction.

Our high values of modeled $A_V$ are in very good agreement with other
indicators of extinction over the M17-SW field, i.e, away from the central
region of NGC 6618.
For example, determinations based on reddening toward individual M17-SW stellar
sources have presented the following values:
$A_V\sim40$ mag for the optically thickest regions, as well as $A_V>25$ mag
for the molecular cloud core based on the $K_s$-band luminosity function
\citep{Jiang2002};
a range of $14\lesssim A_V\lesssim30$ mag for 55 stars within 0.7 pc of the
KW object, as well as $A_V$ of 24 and 30 mag toward the two components of
the KW object itself \citep{Chini2004};
and $A_V>30$ mag for a large number of sources along the reddening vector
\citep{Hoffmeister2008}.
Furthermore, even higher extinction estimates have been derived from far-IR
and mm-wave observations that dissect the entire molecular cloud surrounding
our field of view toward M17-SW.
For example, \cite{Gatley1979} found $A_V\sim100$ mag through the core of the
molecular cloud;
\cite{Thronson1983} estimated $A_V\leq200$ mag at the peak of $^{13}$CO
emission;
\cite{Keene1985} found $A_V\approx100$ mag based on $^{13}$CO data;
and finally, \cite{Wilson2003} decomposed cloud B into individual clumps with
inferred $N$(H$_2)=(1.9$---$10.4)\times10^{22}$ cm$^{-2}$ for clumps
B25, B27, B29, B32, and B34, which are the ones that are either partially or
fully overlapped by our \textit{Spitzer} field of view.
Focusing on B27 and B29, the two $^{13}$CO clumps wholly enclosed inside
our field, and
employing the relationship $A_V=2\times N$(H$_2)/1.8\times10^{21}$ mag,
values of $A_V$ = 21 and 29 mag can be inferred through these two clumps in
cloud B.
We consider such a consistent picture as a confirmation that PDR modeling may
be employed for reliable mapping of $A_V$ values toward H$_2$ emission obscured
by dust.

Although we find here very good agreement between PDR-modeled $A_V$
values and values that have been derived by other means,
it is important to cross-check our method toward another well-studied PDR,
owing to the novelty involved.
NGC 2023 is such a PDR, toward which a previously successful
PDR modeling has been achieved with a fixed value of $A_V$
\citep[][and references therein]{Sheffer2011}.
We performed the parameterized-$A_V$ test successfully, resulting in model fits
similar to those
previously obtained, as well as in $A_V$ output consistent with results
from other studies.
Whereas our previous modeling employed a fixed value of $A_K=0.5$ mag,
or equivalently, $A_V=4.6$ mag, the new test returns a field-wide median
of $A_V=8$ mag.
Our PDR modeling is again indicating a positive radial gradient of
extinction across the field, starting with much lower values of $A_V$ in the
region between the SR and the exciting star, HD 37903, and reaching
$A_V\sim20$ mag on the other side of the SR, or deeper into the dense
molecular cloud.
In this case, as is the case with M17-SW, the larger PDR-modeled $A_V$ values
are consistent with previous studies.
\cite{DePoy1990} concluded that over small scales, the extinction toward
stars in NGC 2023 varies over 0--10 mag.
Furthermore, the total extinction through the molecular cloud is expected
to be $\geq$25 mag \citep{DePoy1990}.
Over the SR, the range of $A_V$ values is $10\pm5$ mag, where the inferred
semi-amplitude of the range is comparable to the level of uncertainty
in the visual extinction determinations toward heavily obscured targets
\citep[e.g.,][]{Nielbock2008}.
For M17-SW we find a dispersion of $\pm4$ mag along the fitted gradient.

\section{Visualization of the LOS Dimension}

\begin{figure}
\epsscale{1.2}
\plottwo{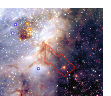}{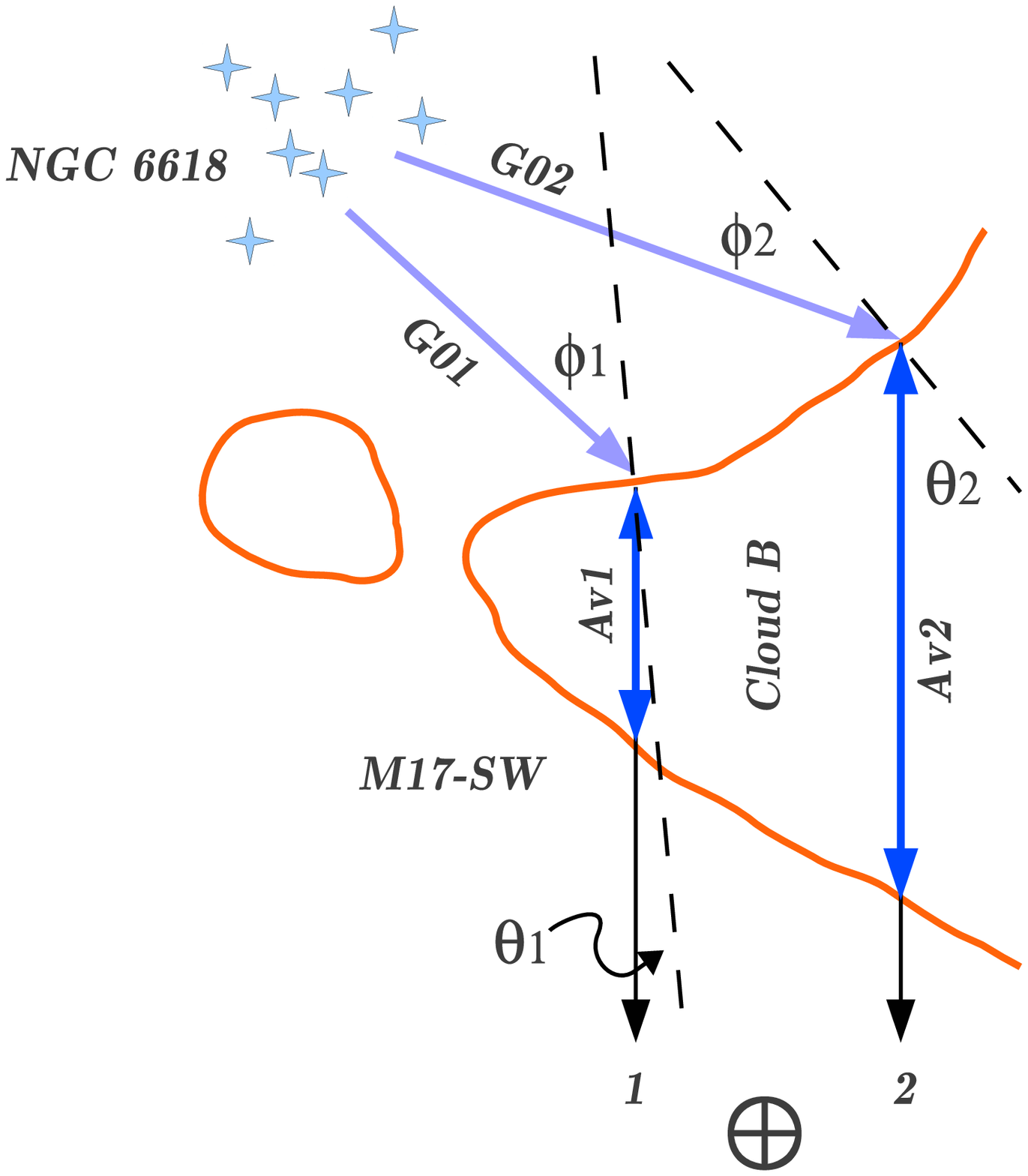}
\caption{Left panel shows the region of IRS data coverage
over a JHK image from Figure 1 of \cite{Jiang2002}.
It can be seen that our \textit{Spitzer} observations (red border) overlap
a dusty ridge of foreground gas, which is obscuring a background PDR
and is producing an extinction with a positive $A_V$
radial gradient.
Right panel presents a schematic interpretation of the 3-D configuration
of M17-SW in terms of a PDR surface on the back side of Cloud B.
Varying distance from NGC 6618 controls the level of incident FUV radiation,
$G_0$, whereas varying thickness of obscuring material controls the level of
visual extinction, $A_V$, in our direction ($\earth$).
Gradual variations in $\theta$ and $\phi$ may explain the behavior of
modeled $f_{\rm eff}$ in terms of changing PDR surface orientation relative
to both NGC 6618 and our LOS.}
\label{fig:fig3}
\end{figure}

A visually stunning depiction of the radially increasing $A_V$ field is
provided in the left panel of Figure~\ref{fig:fig3}, which is based
on Figure 1 from \cite{Jiang2002}.
Our basic assumption is that the observed H$_2$ emission originates on
a cloud surface facing away from our direction and basking in the FUV
starshine of NGC 6618.
We thus view the obscured back side of the PDR surface through the bulk
of cloud B, as measured by our modeled $A_V$ extinction values along the
LOS, see right panel of Figure~\ref{fig:fig3}.
Any \ion{C}{2} or high-$J$ CO line emission \citep[e.g.,][]{Perez2012} would
also be observed from the far side of the intervening molecular cloud and its
interpretation should account for the face-on geometry and intervening
cold gas layer.

The last panel of Figure~\ref{fig:fig2} showed that $f_{\rm eff}<1$, which
means that $\phi>\theta$ under the assumption of $f_{\rm B}=1$, and therefore
a shallow grazing angle for the FUV influx from NGC 6618.
We further assume that the cos($\phi$)/cos($\theta$) curve is
defined by three continuous segments of the PDR surface and not by
local variations over pixel-sized scales.
Each construction of
\clearpage
\clearpage
\noindent a PDR surface begins at any arbitrarily chosen location
along the LOS, and proceeds by
forcing neighboring segments to be smoothly
connected in order to avoid the introduction of FUV shadowing between segments.

\begin{figure}
\epsscale{1.0}
\plotone{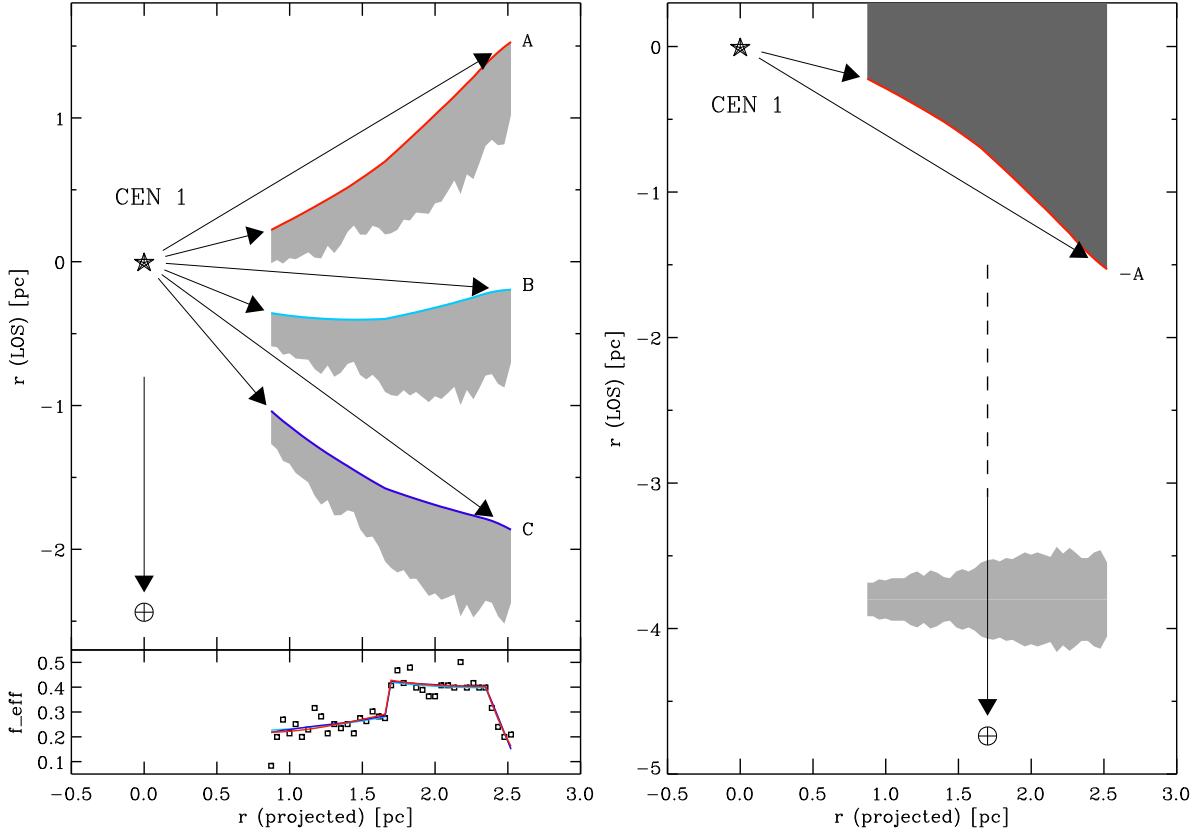}
\caption{Top left panel presents three possible LOS configurations
of the PDR surface, all possessing the same
$f_{\rm eff}=\cos(\phi)/\cos(\theta)$ curves.
Bottom left panel shows such overlapping $f_{\rm eff}$ curves in comparison
with modeled values (squares) taken from Figure~\ref{fig:fig2}.
The (colorized) PDR layer (of thickness $\leq$ 10$^{-4}$ pc) is assumed to
constitute the back side
of an obscuring cloud as viewed along our LOS ($\earth$).
Clouds are projected onto an (X, Y) plane that includes the observer,
the FUV source ($\bigstar$ at the location of CEN 1), and the radial axis of
the mapped PDR area.
The degeneracy in LOS cloud positions can be extended to include
their reflection across the abscissa, as shown in the right panel.
Thus PDR surface `$-$A' is characterized by the same $f_{\rm eff}$ curve as PDR
surface `A'.
In this configuration, however, the intervening extinction toward the PDR
arises in an isolated cloud at an unspecifiable position along the LOS.
The illustrated cloud thickness along the LOS has been converted from modeled
$A_V$ values by employing the arbitrary $n_{\rm H}=4\times10^4$ cm$^{-3}$.}
\label{fig:fig4}
\end{figure}

Figure~\ref{fig:fig4} presents four examples of PDR surfaces that share the
same curve of $f_{\rm eff}$ values.
The first three surfaces, shown in the left panel, belong to our basic
assumption that the PDR is viewed from its back side.
Each derived concave surface is consistent with the general 3-D configuration
of a bowl-shaped interface originally suggested by \cite{Meixner1992},
see also Figure 2 of \cite{Pellegrini2006}.
The right panel show an alternate configuration in which the PDR is viewed
from its front side and is physically separated
from the layer of foreground extinction.
Although the cos($\phi$)/cos($\theta$) curve fixes the relative
curvature of the surface, and thus provides a `3-D view' of the PDR layer
along the LOS, both the absolute location along the LOS
and the state of reflection about the abscissa are, unfortunately, degenerate.

Future studies could incorporate additional clues
for a more robust characterization of the location and the
reflection of the PDR surface.
For example, our initial tests involving a freely variable $G_0$ show that
its values start to drop significantly below $G_0^{\rm pr}$ values half way
along the mapped radial distance from NGC 6618.
Such a behavior may indicate that the line of sight is probing a more
extreme case of geometry than derived here.
Among the PDR surfaces depicted in Figure~\ref{fig:fig4}, only `A', and
therefore `$-$A', affect $G_0$ values in a similar
fashion, owing to their increasing deviation from the plane of projection.
On the other hand, the role of pixel-to-pixel variations, such
as those seen in modeling output, remains to be evaluated in terms of
orientation and mutual shadowing among individual clumps.

\section{Summary}

1. \textit{Spitzer}-IRS spectra were obtained and employed in our quest to
measure the intensity of H$_2$ emission lines toward the PDR M17-SW.

2. Following conversion of H$_2$ intensity into column density, we employed
our PDR code in order to map physical quantities as a function of distance
from the source of FUV radiation. 

3. We introduced $A_V$ as a free parameter into the PDR code, and subsequently
successfully derived radial mapping of the extinction suffered by H$_2$ lines.

4. Our analysis of the data-to-model column density ratio in terms of
$f_{\rm eff}=\cos(\phi)/\cos(\theta)$ provided a `3-D view' of the
line-of-sight structure of the PDR surface, showing it to be
\clearpage
\clearpage
\noindent globally curved, thus confirming the suggested description of
a bowl-shaped PDR in M17-SW.

\begin{deluxetable}{ccccc}
\tablecolumns{5}
\tablewidth{0pt}
\tabletypesize{\footnotesize}
\tablecaption{Log of \textit{Spitzer} Observations of M17-SW}
\tablehead{\colhead{AOR} &\colhead{Date} &\colhead{$\alpha$(J2000)} &\colhead{$\delta$(J2000)} &\colhead{Exposures} \\
 & &(deg) &(deg) &(steps $\times$ cycles $\times$ s) }
\startdata
\multicolumn{5}{c}{Module SH: Slit = 4$\farcs$7 $\times$ 11$\farcs$3, 2$\farcs$26/pixel} \\
\cline{1-5}
11543296  &2004/10/02  &275.11458   &$-$16.19700    &48 $\times$ 1 $\times$ 6.3  \\
11543552  &2005/04/23  &275.11125   &$-$16.20081    &48 $\times$ 1 $\times$ 6.3  \\
11543808  &2005/04/23  &275.10792   &$-$16.20461    &48 $\times$ 1 $\times$ 6.3  \\
11544064  &2005/04/17  &275.10458   &$-$16.20842    &48 $\times$ 1 $\times$ 6.3  \\
11544320  &2005/04/17  &275.10125   &$-$16.21222    &48 $\times$ 1 $\times$ 6.3  \\
11544576  &2005/04/17  &275.09792   &$-$16.21603    &48 $\times$ 1 $\times$ 6.3  \\
11544832  &2005/04/17  &275.08958   &$-$16.21572    &48 $\times$ 1 $\times$ 6.3  \\
11545088  &2005/04/17  &275.08625   &$-$16.21953    &48 $\times$ 1 $\times$ 6.3  \\
11545344  &2005/04/17  &275.08292   &$-$16.22333    &48 $\times$ 1 $\times$ 6.3  \\
11545600  &2005/04/17  &275.07958   &$-$16.22714    &48 $\times$ 1 $\times$ 6.3  \\
11545856  &2005/04/17  &275.07625   &$-$16.23094    &48 $\times$ 1 $\times$ 6.3  \\
\\
\multicolumn{5}{c}{Module LH: Slit = 11$\farcs$1 $\times$ 22$\farcs$3, 4$\farcs$46/pixel} \\
\cline{1-5}
11546112  &2005/04/23  &275.11458   &$-$16.19700    &30 $\times$ 1 $\times$ 6.3  \\
11546368  &2005/04/17  &275.10121   &$-$16.20969    &30 $\times$ 1 $\times$ 6.3  \\
11546624  &2005/04/17  &275.08563   &$-$16.22011    &30 $\times$ 1 $\times$ 6.3  \\
11546880  &2005/04/17  &275.07225   &$-$16.23281    &30 $\times$ 1 $\times$ 6.3  \\
\\
\multicolumn{5}{c}{Module SL: Slit = 3$\farcs$7 $\times$ 57$\arcsec$, 1$\farcs$85/pixel} \\
\cline{1-5}
17976320  &2007/10/12  &275.07213   &$-$16.22614    &30 $\times$ 8 $\times$ 14.7  \\
17976576  &2007/09/30  &275.08188   &$-$16.22347    &30 $\times$ 8 $\times$ 14.7  \\
17977344  &2007/09/30  &275.09183   &$-$16.21867    &30 $\times$ 8 $\times$ 14.7  \\
17977600  &2007/09/30  &275.10083   &$-$16.20853    &30 $\times$ 10 $\times$ 6.3  \\
\enddata
\end{deluxetable}


\begin{thebibliography}{}

\bibitem[Beetz et al.(1976)]{Beetz1976} Beetz, M., Els\"{a}sser, H., Poulakos, C., \& Weinberger, R. 1976, \aap, 50, 41
\bibitem[Brogan \& Troland(2001)]{Brogan2001} Brogan, C. L., \& Troland, T. H. 2001, \apj, 560, 821 
\bibitem[Broos et al.(2007)]{Broos2007} Broos, P. S., Feigelson, E. D., Townsley, L. K., Getman, K. V., Wang, J., Garmire, G. P., Jiang, Z., \& Tsuboi, Y. 2007, \apjs, 169, 353
\bibitem[Chini et al.(1980)]{Chini1980} Chini, R., Els\"{a}sser, H., \& Neckel, Th. 1980, \aap, 91, 186
\bibitem[Chini et al.(2004)]{Chini2004} Chini, R., Hoffmeister, V. H., K\"{a}mpgen, K., Kimeswenger, S., Nielbock, M., \& Siebenmorgen, R. 2004, \aap, 427, 849
\bibitem[DePoy et al.(1990)]{DePoy1990} DePoy, D. L., Lada, E. A., Gatley, I., \& Probst, R. 1990, \apj, 356, L55
\bibitem[Dickel(1968)]{Dickel1968} Dickel, H. R. 1968, \apj, 152, 651
\bibitem[Felli et al.(1984)]{Felli1984} Felli, M., Churchwell, E., \& Massi, M. 1984, \aap, 136, 53
\bibitem[Gatley et al.(1979)]{Gatley1979} Gatley, I., Becklin, E. E., Sellgren, K., \& Werner, M. W. 1979, \apj, 233, 575
\bibitem[Habing(1968)]{Habing1968} Habing, H. J. 1968, Bull. Astron. Inst. Netherlands, 19, 421
\bibitem[Hanson et al.(1997)]{Hanson1997} Hanson, M. M., Howarth, I. D., \& Conti, P. S. 1997, \apj, 489, 698
\bibitem[Hoffmeister et al.(2008)]{Hoffmeister2008} Hoffmeister, V. H., Chini, R., Scheyda, C. M., Schulze, D., Watermann, R, N\"{u}rnberger, D., \& Vogt, N. 2003, \apj, 686, 310
\bibitem[Jiang et al.(2002)]{Jiang2002} Jiang, Z., Yao, Y., Yang, J. et al. 2002, \apj, 577, 245
\bibitem[Kaufman et al.(2006)]{Kaufman2006} Kaufman, M. J., Wolfire, M. G., \& Hollenbach, D. J. 2006, \apj, 644, 283
\bibitem[Keene et al.(1985)]{Keene1985} Keene, J., Blake, G. A., Phillips, T. G., Huggins, P. J., \& Beichman, C. A. 1985, \apj, 299, 967
\bibitem[Kleinmann \& Wright(1973)]{Kleinmann1973} Kleinmann, D. E., \& Wright, E. L. 1973, \apj, 185, L133
\bibitem[Mathis(1990)]{Mathis1990} Mathis, J. S. 1990 \araa, 28, 37
\bibitem[Meixner et al.(1992)]{Meixner1992} Meixner, M., Haas, M. R., Tielens, A. G. G. M., Erickson, E. F., \& Werner, M. 1992, \apj, 390, 499
\bibitem[Meixner and Tielens(1993)]{Meixner1993} Meixner, M., \& Tielens, A. G. G. M. \apj, 405, 216
\bibitem[Nielbock et al.(2008)]{Nielbock2008} Nielbock, M., Chini, R., Hoffmeister, V. H., N\"{u}rnberger, D. E. A., Scheyda, C. M.,  \& Steinacker, J. 2008, \mnras, 388, 1031
\bibitem[Parravano et al.(2003)]{Parravano2003} Parravano, A., Hollenbach, D. J., \& McKee, C. F. 2003, \apj, 584, 797
\bibitem[Pellegrini et al.(2006)]{Pellegrini2006} Pellegrini, E. W., Baldwin, J. A., Brogan, C. L. et al. 2006, \apj, 658, 1119
\bibitem[P\'{e}rez-Beaupuits et al.(2012)]{Perez2012} P\'{e}rez-Beaupuits, J. P., Wiesemeyer, H., Ossenkopf, V., Stutzki, J., G\"{u}sten, R., Simon, R., H\"{u}bers, H. W., \& Ricken, O. 2012, \aap, 542, L13
\bibitem[Povich et al.(2007)]{Povich2007} Povich, M. S., Sone, J. M., Churchwell, E., et al. 2007, \apj, 660, 346
\bibitem[Povich et al.(2009)]{Povich2009} Povich, M. S., Churchwell, E., Bieging, J. H. et al. 2009, \apj, 696, 1278
\bibitem[Sheffer et al.(2011)]{Sheffer2011} Sheffer, Y., Wolfire, M. G., Hollenbach, D. J., Kaufman, M. J., \& Cordier, M. 2011, \apj, 741, 45
\bibitem[Smith et al.(2007)]{Smith2007} Smith, J. D. T., Armus, L., Dale, D. A., et al. 2007, \pasp, 119, 1133
\bibitem[Stutzki et al.(1988)]{Stutzki1988} Stutzki, J., Stacey, G. J., Genzel, R., Harris, A. I., Jaffe, D. T., \& Lugten, J. B. 1988, \apj, 332, 379
\bibitem[Stutzki \& G\"{u}sten(1990)]{Stutzki1990} Stutzki, J., \& G\"{u}sten, R. 1990, \apj, 356, 513
\bibitem[Thronson \& Lada(1983)]{Thronson1983} Thronson, H. A., \& Lada, C. J. 1983, \apj, 269, 175
\bibitem[Tokunaga \& Thompson(1979)]{Tokunaga1979} Tokunaga, A. T., \& Thompson, R. I. 1979, \apj, 229, 583
\bibitem[Wilson et al.(2003)]{Wilson2003} Wilson, T. L., Hanson, M. M., \& Muders, D. 2003, \apj, 590, 895
\bibitem[Xu et al.(2011)]{Xu2011} Xu, Y., Moscadelli, L., Reid, M. J., Menten, K. M., Zhang, B., Zheng, X. W., \& Brunthaler, A. 2011, \apj, 733, 25

\end{thebibliography}
\end{document}